\documentclass[11.5pt, letterpaper]{article}
\usepackage[hmargin=5em,vmargin=7em, bmargin=5em,footskip=3em]{geometry}
\usepackage{times}
\usepackage{amsmath}
\usepackage{graphicx}
\usepackage{natbib}
\usepackage{tikz}
\usepackage{lipsum}
\usepackage{hyperref}
\usepackage{mathtools}
\usepackage{placeins}
\usepackage{gensymb}
\usepackage{amssymb}
\usepackage{authblk}
\usepackage{array}
\usepackage{etoolbox}
\usepackage{boxhandler}
\usepackage{listing}
\usetikzlibrary{positioning}
\usepackage{lineno}

\usepackage{algorithm}
\usepackage{todonotes}
\presetkeys{todonotes}{color=blue!30}{}
\usepackage[noend]{algpseudocode}
\usepackage[font=scriptsize]{caption}

\makeatletter
\def\@footnotecolor{black}
\define@key{Hyp}{footnotecolor}{%
 \HyColor@HyperrefColor{#1}\@footnotecolor%
}
\def\@footnotemark{%
    \leavevmode
    \ifhmode\edef\@x@sf{\the\spacefactor}\nobreak\fi
    \stepcounter{Hfootnote}%
    \global\let\Hy@saved@currentHref\@currentHref
    \hyper@makecurrent{Hfootnote}%
    \global\let\Hy@footnote@currentHref\@currentHref
    \global\let\@currentHref\Hy@saved@currentHref
    \hyper@linkstart{footnote}{\Hy@footnote@currentHref}%
    \@makefnmark
    \hyper@linkend
    \ifhmode\spacefactor\@x@sf\fi
    \relax
  }%
\makeatother
\hypersetup{
    colorlinks = true,
    urlcolor=[rgb]{0.1, 0.1, 0.8},
    menucolor = black,
    citecolor = black,
    footnotecolor = black,
    linkcolor = black,
    backref = true
}
\usetikzlibrary{bayesnet}

\newcommand{\ParamAB}{\pi_{AB}}
\newcommand{\ParamBA}{\pi_{BA}}

\newcommand{\ParamAA}{\pi_{AA}}
\newcommand{\ParamBB}{\pi_{BB}}

\newcommand{\EstParamAB}{\widehat{\pi}_{AB}}
\newcommand{\EstParamBA}{\widehat{\pi}_{BA}}

\newcommand{\EstIntAB}{\widehat{I}_{AB}}
\newcommand{\EstIntBA}{\widehat{I}_{BA}}

\newcommand{\AProgram}{\textnormal{A-Program}}
\newcommand{\BProgram}{\textnormal{B-Program}}
\newcommand{\AToB}{\textnormal{A-To-B}}
\newcommand{\BToA}{\textnormal{B-To-A}}
\newcommand{\XToY}{\textnormal{X-To-Y}}

\newcommand{\AToBRep}{\textnormal{A-To-B}_{R}}
\newcommand{\BToARep}{\textnormal{B-To-A}_{R}}
\newcommand{\XToYAct}{\textnormal{X-To-Y}_{A}}
\newcommand{\XToYRep}{\textnormal{X-To-Y}_{R}}

\newcommand{\history}{\mathsf{history}}

\newcommand{\ASensor}{\textnormal{A-Sensor}}

\newcommand{\ARecorder}{\textnormal{A-Recorder}}

\newcommand{\BSensor}{\textnormal{B-Sensor}}
\newcommand{\BRecorder}{\textnormal{B-Recorder}}

\newcommand{\XSensor}{\textnormal{X-Sensor}}

\newcommand{\YRecorder}{\textnormal{Y-Recorder}}

\newcommand{\Bern}{\mathrm{Bern}}
\newcommand{\BetaDist}{\mathrm{Beta}}
\newcommand{\Bino}{\mathrm{Bin}}

\begin{document}


\title{Embodying probabilistic inference in biochemical circuits}

\author[1,2]{Yarden Katz\thanks{Correspondence to: yarden@hms.harvard.edu}}
\author[1]{Michael Springer}
\author[1]{Walter Fontana}
\affil[1]{\small Dept. of Systems Biology, Harvard Medical School, Boston, MA}
\affil[2]{\small Berkman Klein Center for Internet \& Society, Harvard University, Cambridge, MA}

\date{\small \today}

\maketitle

\begin{abstract}
Probabilistic inference provides a language for describing how
organisms may learn from and adapt to their environment. The
computations needed to implement probabilistic inference often require
specific representations, akin to having the suitable data structures for
implementing certain algorithms in computer programming. Yet it is unclear how such
representations can be instantiated in the stochastic,
parallel-running biochemical machinery found in
cells (such as single-celled organisms). Here, we show how
representations for supporting inference in Markov models can be embodied in
cellular circuits, by combining a
concentration-dependent scheme for encoding probabilities with a
mechanism for directional counting. We show how
the logic of protein production and degradation constrains the
computation we set out to implement. We argue that this process by
which an abstract computation is shaped by its biochemical
realization strikes a compromise between ``rationalistic''
information-processing perspectives and alternative approaches that
emphasize embodiment.
\end{abstract}

%
%


%
%
%
%
%
%

\maketitle
%
%


\section*{Introduction}

\begin{quote}
\small 
\textit{``Quality, light, color, depth, which are there before us, are there only because they awaken an echo in our bodies and because the body welcomes them. Things have an internal equivalent in me; they arouse in me a carnal equivalent of their presence.'}' 
                                \begin{flushright}---Maurice Merleau-Ponty, \textit{Eye and Mind} (1964)
\end{flushright}\end{quote}

Probabilistic inference has emerged as a useful language for
describing various aspects of cognition
\citep{Knill2004BayesianBrain,Griffiths2010ProbabilisticModels,LakeUllman2017BuildingMachines}. In
this line of work, people (and other animals) are modeled as if they represent probabilistic
models of their world and use inference in these models to guide
perception and action. While much of the work on biological realizations of probabilistic inference has focused on neural circuits
\citep{Knill2004BayesianBrain,Gallistel2013BeyondHebb}, single-celled
organisms also face dynamic and uncertain environments that necessitate a
coordinated response by the organism. This situation can be framed in cognitive terms
such as decision-making and inference, and indeed, frameworks from cognitive science
have been applied to the biology of both single
cells and microbial populations
\citep{Bray1990PDPCells,Jacob2004BacterialSocial,Fernando2009HebbianCell,Lyon2015CognitiveCell,Katz2016PeerJ,BaluskaLevin2016OnHavingNoHead}. Yet
probabilistic inference is challenging to realize using cellular
biochemistry. In a single cell, for instance, protein
interactions take place stochastically and in parallel, without the
direction of a centralized ``conductor'' that imposes
sequential order. How can such biochemical substrates support probabilistic
inference? 

In theory, Chemical Reaction Networks (CRNs), like sufficiently elaborate
neural networks, can approximate any computable function \citep{Hjelmfelt1991NeuralNetworksCRN,DBLP:conf/cmsb/FagesGBP17} and therefore can
perform inference. For instance, \citep{Hjelmfelt1991NeuralNetworksCRN} showed how to implement
McCulloch-Pitts neurons, and thus arbitrary Boolean circuits, in
CRNs. Probabilistic computation with CRNs has also been explored:
\cite{NappAdamsNIPS2013} showed how to construct a chemical reaction network whose
steady-state encodes the marginals of joint probability distributions
in a class of probabilistic models known as factor graphs. As
\citep{NappAdamsNIPS2013} point out, enzyme-free DNA strand displacement
\citep{Soloveichik2010StrandDisplacement} can serve as the
physical basis of such CRNs. However,
while CRNs can encode these probabilistic computations, only a
subset of CRNs can be plausibly instantiated using
cellular signaling (and the nucleic acids-based implementation of CRNs
\citep{Soloveichik2010StrandDisplacement} is arguably not a
plausible intracellular mechanism). Also, the reliance on
steady-states does not capture many of the complexities of a cellular context. A cell
responds to perturbations from its environment in real-time, and does so
in the presence of considerable stochastic fluctuations in gene expression
\citep{Raj2008NoisyGeneReview,Balazsi2011CellularNoise} (such
fluctuations are not taken into account in the steady-state framework of
\citep{NappAdamsNIPS2013}). This raises the question of
what biochemical circuits could theoretically realize inference in the
dynamic and stochastic environment in which cells live.

Several studies looked for biochemical circuits within single-celled
organisms that could exhibit
``predictive'' behavior in a dynamic environment, through both laboratory and \textit{in silico}
evolution \citep{TagkopoulosScience2008,MitchellNature2009}. As argued
in \citep{McGregor2012EvolutionAssociative}, however, in these
experiments, cells (or the biochemical circuits that were evolved
\textit{in silico}) do
not ``learn'' from the environment but rather are selected to fit
it. Evolving a circuit that confers differential reproductive increase 
in a specific environment is distinct from constructing a biochemical
circuit that learns from the environment (which can take place on timescales faster than those
required for mutation and selection). A variety of epigenetic
mechanisms for retaining memories of the environment have been
described
\citep{Jablonka2009EpigeneticInheritance}
which could potentially support faster learned responses. These
epigenetic mechanisms have been shown to support memory in single-celled
organisms \citep{WolfMemory2008,Lambert2014,Stockwell2015}. For
example, \citep{WolfMemory2008} showed that \textit{B. subtilis}
populations exposed to distinct environmental perturbations and
then grown in the same environment can be distinguished (by
their gene expression profiles) for as long as 24 hours after the
perturbation (suggesting cells retain ``memory'' of their
environment's history). 

These experiments suggest that epigenetic mechanisms could potentially be used to implement learning mechanisms in
cells, and this possibility has been explored theoretically. \citep{Fernando2009HebbianCell} proposed a biochemical
circuit that implements simple conditioning through Hebb's rule, while
\citep{McGregor2012EvolutionAssociative} evolved \textit{in silico} a biochemical circuit
for associative learning. In previous work \citep{Katz2016PeerJ}, we have framed the problem of adaptation to
changing environments as probabilistic inference, a framing in which
single-celled organisms estimate the dynamics of their stochastic
environment and use these estimates to epigenetically regulate their metabolic state. We have shown
in simulation that adaptive strategies based on inference can lead to increased
population growth in changing environments (particularly in complex ``meta-changing'' environments whose dynamics change according to
latent stochastic states \citep{Katz2016PeerJ}). 

In this paper, we focus on constructing, starting with a probabilistic model,
a complete biochemical circuit that realizes inference in this model. To
construct such a circuit, we would need to find biochemical
representations of the environment that can support inference and be usefully connected to other pathways in the cell. 
Some cognitive scientists have argued that in order to be interoperable
with multiple cognitive processes, representations have to be ``computationally accessible''
\citep{Gallistel2008RepresentationReview}. For computational
accessibility, it is not sufficient to have the system ``contain" the
information in its states in an information-theoretic sense (as was
shown for memory in \textit{B. subtilis} \citep{WolfMemory2008}). Rather, the
information has to be encoded in a form that is usable by other
computational processes that are available to the system. We aim to construct circuits that use signaling components
that might plausibly reside in cells, that could withstand stochastic
fluctuations in component levels, and that use molecular representations that
are accessible (i.e., composable with other molecular circuits).

This paper is organized as follows. We first introduce the use of
probabilistic inference to anticipate an environment that follows Markovian dynamics, and derive the
computational representations needed to implement inference in real-time. We then describe a
circuit, embodied in proteins, that can perform these operations using a scheme where
probabilities are encoded as concentrations. We outline several
properties of the circuit (including its failure modes) and show how
these arise as a consequence of the logic of protein production. We close
with a discussion of some implications of our analysis for understanding biological systems in information-processing terms.

\section*{Analysis} 

\subsection*{Real-time probabilistic inference in a changing environment}

Probabilistic models have been used to capture many features of
dynamic and uncertain environments. One of the simplest 
probabilistic models for modeling change is the discrete-time, finite-state Markov model. In
this model, it is assumed that the state of the environment at time $t$ is dependent
only on the prior states going back to the $t - k$ time point, where $k$ is the order of the model (when $k = 1$, we have a first-order
Markov model). While this Markov assumption is violated by many natural
processes \citep{YuHSMM2010}, it can still serve as a useful
idealization (or null model) in some
contexts. We chose to focus on this simple model in order to
better understand how biochemical circuits that perform
inference can be systematically derived from a probabilistic
model's description.

\begin{figure}[t!]
\centering
\includegraphics{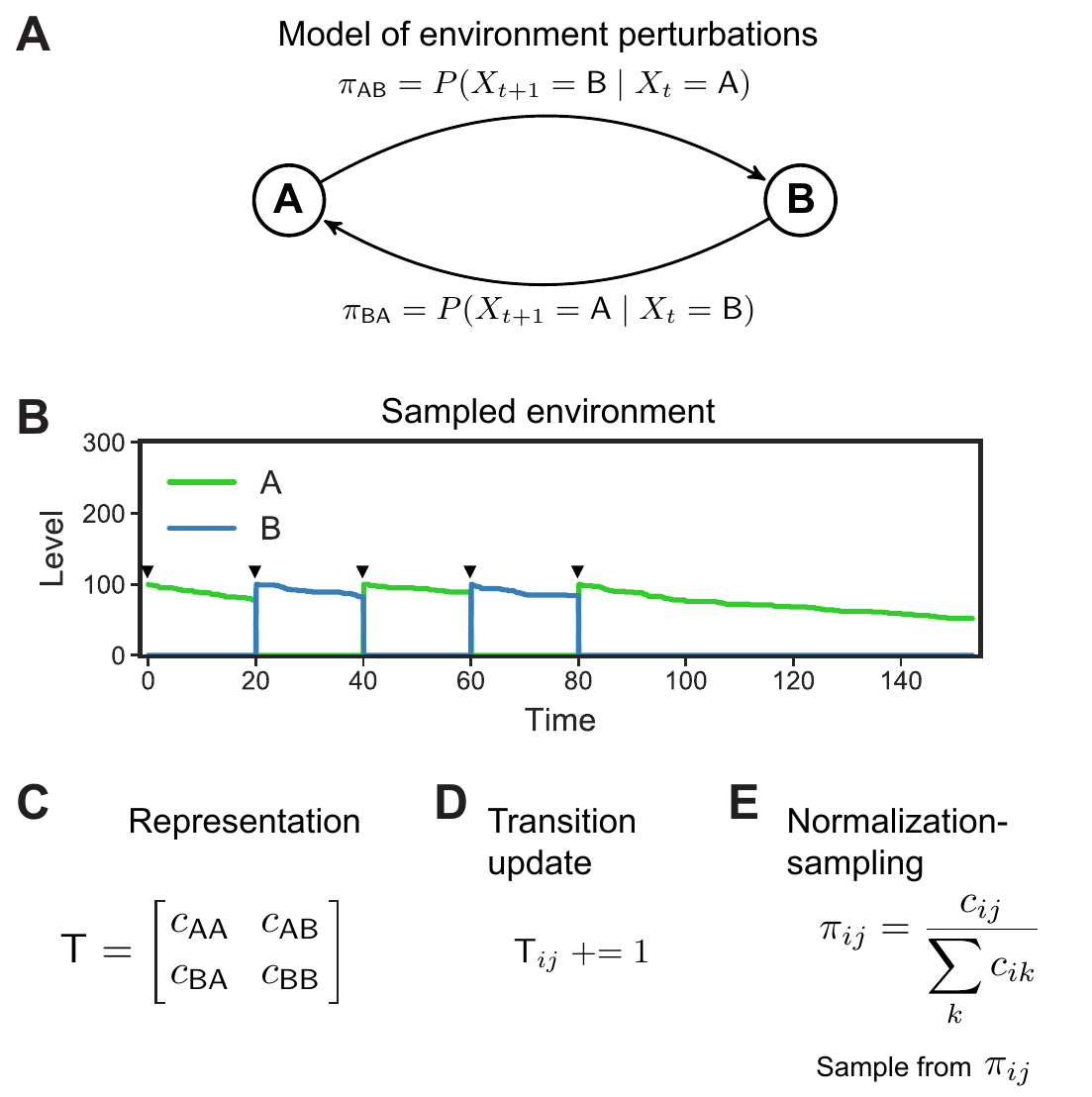}
\caption{Computational representations and operations for inference in Markov
  environments. (A) A discrete-time Markov model that governs switches
  between two states, $A$ and $B$. Model is parameterized by two
  transition probabilities, $\ParamAB$ and $\ParamBA$. (B) An
  environment generated using the model shown in (A). At fixed 20
  time-unit intervals,  100 units of $A$ or $B$ are pulsed in (after
  removing any prior $A$ or $B$) and allowed to degrade. Pulses
  denoted by arrow heads. (C) Transition matrix $T$ that represents the
  sufficient statistics for the model shown in (A). (D) Transition
  operation on $T$, which increments the appropriate counter
  when the environment switches from state $i$ to $j$ (e.g., $A$ to
  $B$). (E) Normalization-sampling operation, which normalizes a row
  in $T$ (converting counts to probabilities) and samples an entry in proportion to its probability.}
\label{figMarkovModel}
\end{figure}

For simplicity, we consider a changing environment that can be in
one of only two states, $A$ or $B$, and in which switches between 
states are driven by a first-order Markov model in discrete time. Such
models are parameterized by two \textit{transition probabilities}: the
probability of switching from $A$ to $B$, denoted $\ParamAB$, and the
probability of switching from $B$ to $A$, $\ParamBA$---as shown in
Fig.~\ref{figMarkovModel}A. This model produces sequences of $A$s and $B$s
that can be used to generate perturbations in a chemical environment (e.g.,
a bioreactor containing microbes where nutrients are flowed in and
out). We used these sequences to add/remove $A$ or $B$ at fixed
time intervals using the
following procedure. At first, a state $X_0 \in \{A, B\}$ is sampled from the
Markov model and then a fixed amount of $X_{0}$ is added to a reactor
which contains our circuit but has no $A$ or $B$ in it. We assume that $A$ and
$B$, which can be thought of as nutrients, are degraded by the circuit
at some constant rate. After a fixed time interval, we sample $X_{t+1}$
from the model given the previous state $X_{t}$. If $X_{t+1} = X_{t}$, no
perturbation is performed, and if $X_{t+1} \neq X_t$, we remove all
$A$ and $B$ present in the reactor and add $X_{t+1}$. An environment
generated by this procedure using a Markov model where $\ParamAB
= \ParamBA = 0.95$ is shown in Fig.~\ref{figMarkovModel}B.

Assuming the environment's perturbations are generated by a Markov model,
probabilistic inference can be used to anticipate the environment's
future states. In particular, it is useful to compute the
probability of encountering $A$ or $B$ next, given past
observations of the environment's states, $\history = \left<X_t,
  X_{t-1}, X_{t-2}, \dots\right>$, where $X_{t}$ corresponds to the
state of the environment at time $t$. In Bayesian terms, anticipation of the environment
often means computing the posterior predictive distribution, $P(X_{t+1} \mid \history)$. This computation is
complicated by the fact that the transition probabilities $\ParamAB$
and $\ParamBA$ are unknown. The standard strategy in such cases is to
place prior probabilities on these parameters and then integrate them
out to calculate, $\int P(X_{t+1} \mid \history, \ParamAB, \ParamBA)
d\ParamAB d\ParamBA$. An alternative and sometimes simpler strategy is to
estimate the unobserved transition probabilities, $\ParamAB$ and
$\ParamBA$, and use these estimates to anticipate the next state. We
can estimate these transition probabilities using Bayesian
inference:
\begin{equation}
\label{posteriorEq}
P(\ParamAB, \ParamBA \mid \history) \propto P(\ParamAB, \ParamBA \mid
\history)P(\ParamAB, \ParamBA)
\end{equation} 


If a mathematically convenient prior distribution is chosen for
$P(\ParamAB, \ParamBA)$, then Eq.~\ref{posteriorEq} can be solved exactly
(see \nameref{Methods}). Using the posterior distribution, we can then obtain estimates of the transition
probabilities, $\EstParamAB, \EstParamBA$, and use these to anticipate
the next state by sampling $X_{t+1} \sim P(X_{t+1} \mid X_t,
\EstParamAB, \EstParamBA)$:

\[ P(X_{t+1} = A \mid X_{t}, \EstParamAB, \EstParamBA) = \begin{cases} 
      1 - \EstParamAB & \textrm{ if } X_t = A\\
      \EstParamBA & \textrm{ if } X_t = B\\
   \end{cases}
\]

An important complication (as pointed out in \citep{Katz2016PeerJ}) is
that organisms act in their environment while it is changing, so the
posterior distribution in Eq.~\ref{posteriorEq} must
be estimated in real-time. However, in our setting, this computation simplifies considerably. For one, the sequence of
observations about the environment, $\history$, does not have to be
stored in full, but can instead be compressed into a matrix $T$ of
transition counts (Fig.~\ref{figMarkovModel}C). Assuming $n$ possible states
of the environment, $T$ is an $n \times n$ matrix whose $i$th row
indicates the number of times the environment switched from the $i$th
state to each of the states. Each row can be further compressed: since
the sum of the $i$th row must equal the total number of transitions
from state $i$ that have been observed, the matrix can be summarized
by $n^2 - n$ entries. In an environment with two states ($A$ or $B$ in
our case), the sufficient statistics are therefore two counts: the
number of times the environment switched from $A$ to $B$, and from
$B$ to $A$ (see \nameref{Methods} for details).

Using these counts as the representation, an elegant algorithm for
computing the posterior distribution in real-time
emerges. As the environment changes, update the relevant counts in
$T$. Then anticipate the next state
given the current state $i$ by: (1) normalizing the $i$th row of $T$\
to convert counts into probabilities, and (2) sampling a state from
that row. This algorithm relies on several computational
operations. First, it is necessary to update the counts by a
\textit{transition update} operation (Fig.~\ref{figMarkovModel}D) that
distinguishes $A \rightarrow B$ transitions from $B \rightarrow A$
transitions. Second, in order to use these counts in downstream
computations, we need to be able to convert the counts into
probabilities by normalization and sample from the
resulting distribution; this can be done via the \textit{normalization-sampling}
operation (Fig.~\ref{figMarkovModel}E). With these three
ingredients---a representation of transition counts in the form of
$T$, a transition update operation that operates on $T$, and a way to
normalize and sample from a row of $T$---the posterior distribution can be used in real-time. 

\begin{figure}[h]
\centering
\includegraphics[scale=0.9]{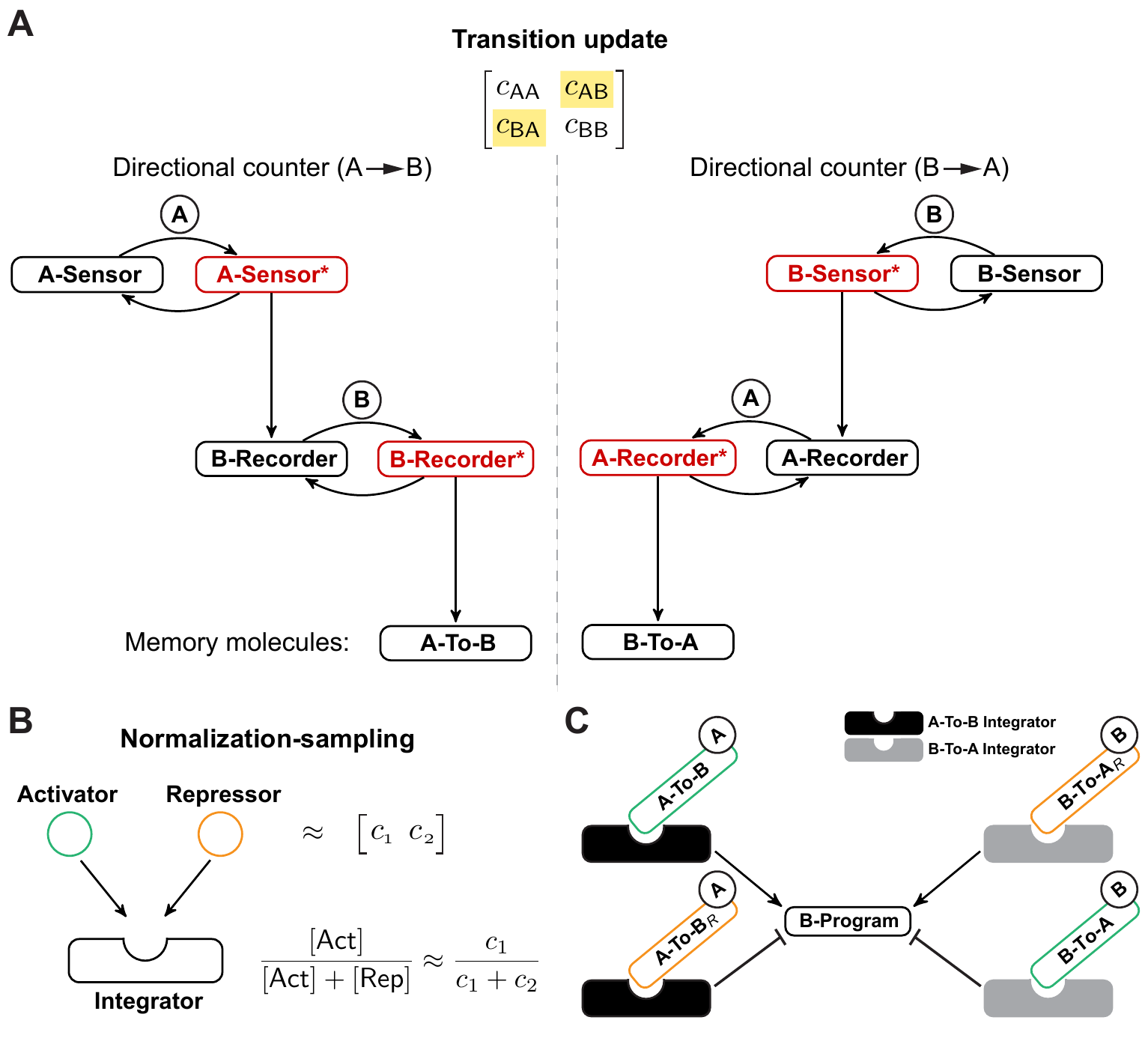}
\caption{Design of a biochemical circuit that performs inference. (A) Biochemical implementation of transition update operation. On
  left, the arm of the circuit that counts $A \rightarrow B$ transitions. The number
  of such transitions ($c_{AA}$ entry in transition matrix) is encoded in the abundance of the $\AToB$ memory
  molecule. On right, symmetric arm that counts $B \rightarrow A$
  transitions ($c_{BA}$ entry in transition matrix), encoded in the
  abundance of the $\BToA$ memory molecule. (B) Normalization-sampling
  operation for converting a row from a transition matrix into
  a probability vector and sampling an element from it. An Activator and 
  Repressor protein each bind a site on an Integrator in mutually
  exclusive fashion. (C) Memory molecules from transition update
  operation combined with normalization-sampling operation to control the
  expression of a molecule associated with $B$ state of environment,
  $\BProgram$. Memory molecules that become activators bound by $A$ or
  $B$ and an integrator are shown in green outline. Repressors shown
  in orange outline; $\AToBRep$ is repressor associated with $\AToB$
  Integrator (similarly for $\BToARep$).}
\label{figCircuit}
\end{figure}

\subsection*{Representing a changing environment using
sensors, recorders and integrators}

To implement probabilistic inference for the model shown
in Fig.~\ref{figMarkovModel}, we need to construct a biochemical
circuit that performs the computational operations outlined in the
previous section. Such a circuit would need a representation of the
transition matrix $T$, which is used to record the number of switches
between relevant states of the environment. Two operations on $T$
would then need to be realized molecularly: (1) the ability to
\textit{count directionally}, i.e., to record when the environment has
switched from state $A$ to $B$ (as opposed to from $B$ to $A$) and
store this information in $T$, and (2) the ability to reference a row
in $T$, normalize it, and sample from the resulting probability
distribution (Fig.~\ref{figMarkovModel}). 

The counts matrix and its associated operations can be
implemented using the scheme shown in
Fig.~\ref{figCircuit}A. Directional counting requires 
counting a potentially unbounded number of switches in the environment. One
approach is to count such transitions using digital counters proposed
in the synthetic biology literature, such as
\citep{Friedland2009SyntheticCounter}, but this counter's capacity is
fixed and scales with the number of genetic components in the circuit---making it
unsuitable for our purpose. We instead represent the number
of transitions using abundances. In this
scheme, each of the different states ($A$ or $B$) has a dedicated
\textit{sensor} that, at any given point in time, can sense the
presence of $A$ or $B$. The environmental states $A$
and $B$ here can be thought of as ligands (e.g., nutrients) that are sensed by receptors (e.g., on
the cell's surface). The transition update operation on $T$ can be
realized as follows. When bound by $A$, the
$A$-sensor triggers the production of another molecule, a
\textit{recorder}. In the presence of $B$, the recorder ($\BRecorder$)
triggers the production of a stable \textit{memory molecule}, $\AToB$, whose
abundance encodes the number of times an $A \rightarrow B$
transition has occurred (Fig.~\ref{figCircuit}A, left). In effect, the
presence of $A$ ``prepares'' the system to record the presence of $B$
(by producing $\BRecorder$). This constitutes one arm of the transition update operation. The second arm of the circuit is symmetric, this time starting with $B$. The $B$-sensor triggers
the production of another recorder molecule ($\ARecorder$), the latter
causing the production of $\BToA$ memory molecules (whose concentration encodes
the number of $B \rightarrow A$ transitions) (Fig.~\ref{figCircuit}A,
right). Note that if we were to examine this
counter circuit design as if it were a naturally occurring signaling pathway, it might seem as though it suffers from
``crosstalk'' between the ``$A$ pathway'' (which starts with sensing
of $A$) and the ``$B$ pathway'' (which starts with the sensing of
$B$). However, this feature is precisely what makes it possible to record the
direction of the switch.\footnote{The physiologist
  H. S. Jennings noted the importance of directional sensing by
  unicellular organisms in his early study, \textit{The
    Behavior of the Lower Organisms} (1906)
  \citep{Jennings1906BehaviorLowerOrganisms}. Jennings
  observed that in
  response to a variety of environmental changes (such as shifts
  in temperature or salt concentration), a Paramecium cell's behavior
  depends upon the direction of change: ``The animal,
  having been subjected to certain conditions, becomes now subjected
  to others, and \textit{it is the transition from one state to
    another that is the cause of the reaction.} This is a fact of
  fundamental significance for understanding the behavior of lower
  organisms. But it is not mere change, taken by itself, that causes
  the reaction, but change \textit{in a certain direction.} [emphasis
  in original.]'' \citep{Jennings1906BehaviorLowerOrganisms}.}

The second operation needed for implementing inference is normalization-sampling
(Fig.~\ref{figCircuit}B). Normalization-sampling requires accessing a row of
$T$ (which contains counts), normalizing its values to obtain probabilities, and then sampling
from the resulting probability distribution. On an ordinary digital computer, using a sequential
programming paradigm, this is straightforward to implement. A
standard algorithm for sampling from a probability vector $\theta$ (in effect, sampling from a multinomial distribution)
is: calculate a cumulative sum up to each element  
$\theta_i$, then iterate through $\theta$, draw a
uniform random number $w \in [0, 1]$ for each entry $\theta_i$, and return
the first $\theta_i$ such that
$\sum_{j=1}^{i}\theta_{j} \geq w$. Inside the cell, however, there is no readily
accessible notion of sequential iteration to support this procedure; a parallelized
mechanism is needed instead. 

We can make use of the fact that concentrations lend themselves to
encoding probabilities (an idea also used in
\citep{NappAdamsNIPS2013}) to derive a non-sequential version of
the same computation. Consider two proteins, an Activator and a
Repressor, that have expression levels $c_1$ and $c_2$, respectively
(Fig.~\ref{figCircuit}B). We can normalize and sample from the
(unnormalized) vector $\theta = [c_1\ c_2]$ by having the
Activator and Repressor bind, in mutually exclusive fashion, a third molecule which we denote 
\textit{Integrator}. If Integrator is not in excess of
Activator and Repressor, then the fraction of Activator-bound Integrator will be proportional to the
concentration of Activator \textit{normalized} by the sum of
concentrations of Activator and Repressor: $\frac{[Act]}{[Act] +
  [Rep]}$ (Fig.~\ref{figCircuit}B). This means that a molecular
interaction that depends on the concentration of Activator-bound
Integrator will occur in proportion to the probability
$\frac{c_1}{c_1 + c_2}$. Note that this does not rely on sequential order; the
interactions between Activator, Repressor, and Integrator can all take place
in parallel. Normalization of protein abundances in this manner is conceptually
similar to the ``divisive normalization'' operation that has been
observed in various neural circuits \citep{CarandiniHeeger2012DivisiveNormalization}.

We can put the transition update and normalization-sampling operations
together to regulate a gene expression program of interest. For example, a protein
associated with the $B$ state, $\BProgram$, can be regulated in a
manner proportional to the probability of encountering $B$, as shown
in Fig.~\ref{figCircuit}C. In this scheme, memory molecules bind integrators in the presence of $A$ or $B$, and the
resulting complex---consisting of memory molecule, integrator and the relevant
signal ($A$ or $B$)---can then function to upregulate or inhibit
$\BProgram$ (Fig.~\ref{figCircuit}C). 

This scheme results in a circuit that uses the posterior over
transition probabilities to regulate the expression of the
environmental state-specific program. Consider the case where the environment is in state $A$. In
anticipation of the $B$ state, it is reasonable to set the level of $\BProgram$
proportional to the probability of switching from $A$ to $B$,
$\ParamAB$. The abundance of the $\AToB$ molecule, which reflects the
number of $A \rightarrow B$ switched experienced so far, contains part
of this information, but has to be converted into a probability. This is
achieved by having the $A$-bound $\AToB$ memory molecule bind an
integrator (shown in black in Fig.~\ref{figCircuit}C), with the resulting
complex functioning as an activator that upregulates
$\BProgram$. Another molecule, $\AToBRep$, when $A$-bound, binds to the same
integrator site as the $A$-bound $\AToB$, turning the complex into a repressor that
downregulates $\BProgram$ (Fig.~\ref{figCircuit}C). This effectively
normalizes $\AToB$ to produce a probability. In the case when the
environment is in state $B$, a symmetric circuit that uses a distinct
integrator (shown in grey in Fig.~\ref{figCircuit}C) enacts the same
logic using the $\BToA$ memory molecule.

\subsection*{Programmatic generation of a biochemical circuit that
  performs inference}

%

\begin{figure}[ht]
\centering
\includegraphics{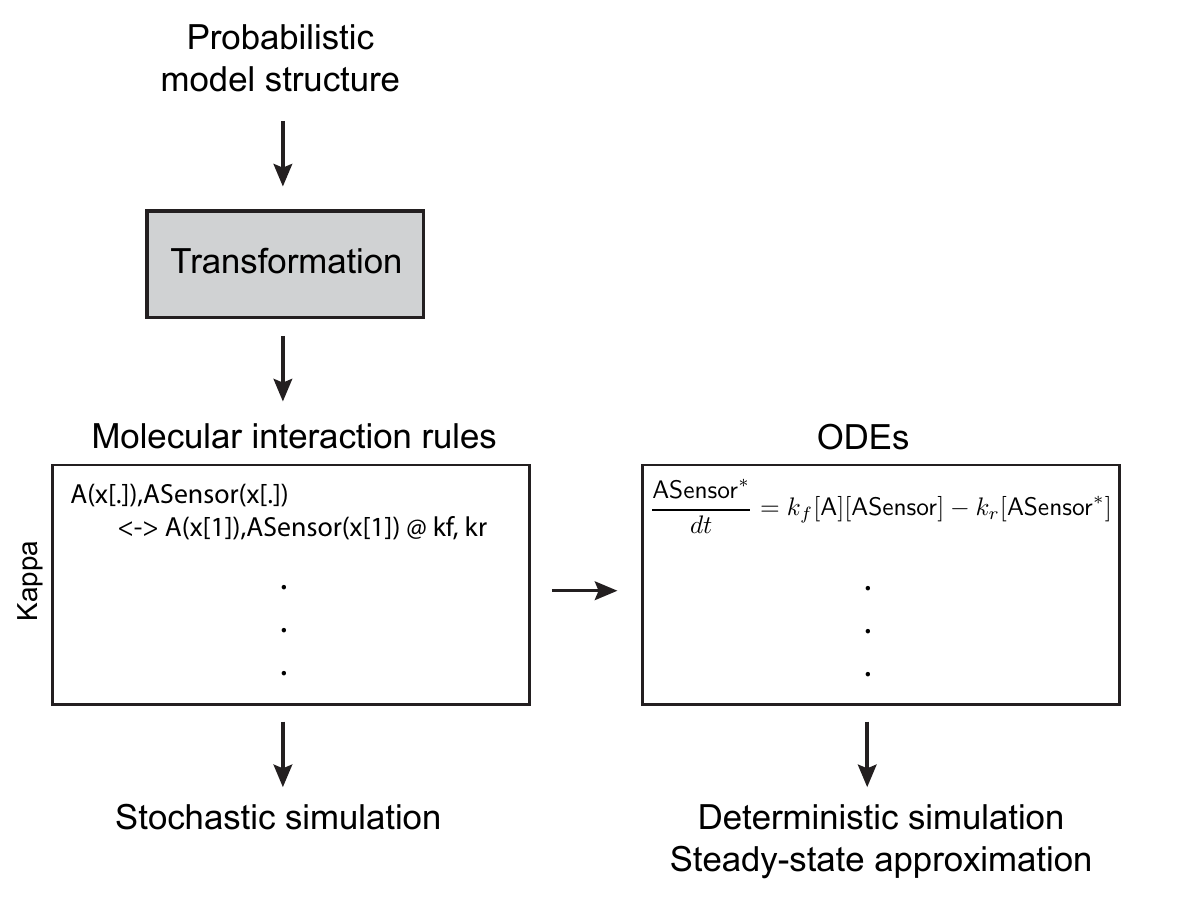}
\caption{Transforming probabilistic model into an executable
  biochemical circuit. Model structure is used to generate
  molecular interactions rules (in Kappa language \citep{Danos2004Formal})
  corresponding to the circuit in Fig.~\ref{figCircuit}, which can
  be used for stochastic simulation (left), and a set of ODEs for
  deterministic simulation and steady-state approximation (right).} 
\label{figProgram}
\end{figure}

Putting the circuit parts described above together, we can now
give a procedure for converting a Markov model into a description
of a biochemical circuit that performs inference in this model. We have implemented a computer
program (in Python) that, taking advantage of the symmetries in the
circuit's design, generates a description of the circuit in the Kappa
language \citep{Danos2004Formal,Danos2007ScalableSimulationRules}. Kappa is a rule-based
language describing the direct molecular interactions among proteins
and simulating these interactions stochastically
(Fig.~\ref{figProgram}). A rule-based representation is particularly convenient for capturing our circuit,
partly because the circuit consists of proteins that play
distinct roles depending on the configuration of their binding
sites (as in Fig.~\ref{figCircuit}D). While we will focus mainly on
stochastic simulation below, Kappa programs can be automatically converted into a set of
ODEs that can be used for deterministic simulation and approximation of
steady-states (Fig.~\ref{figProgram}).\footnote{In general, a Kappa
  program may correspond to an intractably large number of ODEs (due
  to combinatorial explosion of species \citep{Feret2009CoraseGrain}), but this is not an issue for
our circuit which has a relatively small number of components and
binding sites.} The high-level procedure for generating a circuit that
performs inference in a Markov model parameterized by a $2 \times 2$ transition matrix $T$ is as follows (rate constants omitted for clarity):

\begin{enumerate}
\item For the first row in $T$: let $X \in \{A, B\}$ and $Y$ be the other state, $Y \in \{A,
  B\}, Y \neq X$.
\item Generate rules for transition update:
\begin{enumerate}
\item Let $\XSensor$ be a sensor reversibly activated by $X$: $X + \XSensor \xrightleftharpoons{} {\XSensor}^{*}$
\item When active, $\XSensor$ triggers production of $\YRecorder$:
  ${\XSensor}^{*} \longrightarrow {\XSensor}^{*} + \YRecorder$
\item Let $\YRecorder$ be a recorder reversibly activated by $Y$: $Y +
  \YRecorder \xrightleftharpoons{} {\YRecorder}^{*}$
\item When active, $\YRecorder$ triggers production of $\XToY$ memory molecule
\end{enumerate}
\item Generate rules for normalization-sampling:
\begin{enumerate}
\item Let $\XToY$ be reversibly activated by $X$ to form Activator (denoted $Act$):
  $X + \XToY \xrightleftharpoons{} Act$
\item Let $\XToYRep$ be reversibly activated by $X$ to form Repressor (denoted $Rep$): $X + \XToYRep \xrightleftharpoons{} Rep$
\item Let Integrator (denoted $Int$) be bound in mutually exclusive fashion by either
  Activator or Repressor to form the following complexes:
\begin{enumerate}
\item Activator reversibly binds Integrator to form $\XToY$ activation
  complex, $\XToYAct$: $Act + Int \xrightleftharpoons{} \XToYAct$
\item Repressor binds Integrator to form $\XToY$ repression complex,
  $\XToYRep$:
$Rep + Int \xrightleftharpoons{} \XToYRep$
\end{enumerate}
\end{enumerate}
\item Generate rules for regulating state-specific expression
  programs\footnote{Only the rules for $\BProgram$ are shown
    here. Symmetric rules can be generated to regulate $\AProgram$.}:
\begin{enumerate}
\item $\XToYAct$ upregulates $\BProgram$: $\XToYAct \longrightarrow
  \XToYAct + \BProgram$ 
\item $\XToYRep$ downregulates $\BProgram$: $\XToYAct + \BProgram \longrightarrow \XToYAct$
\end{enumerate}
\item Repeat above steps for the remaining row of $T$ (i.e., if the
  above steps were executed with $X = A, Y = B$, now repeat with $X = B, Y = A$).
\end{enumerate}

We generated a Kappa program using this procedure (see
\nameref{Methods} for full program). We chose rate
constants for various steps (e.g., activation of sensors) that were were biophysically plausible (on par with rates of protein
production and post-translational modifications) and assumed that
sensors are constitutively expressed. The levels of integrators and
repressors were set to be roughly equal. In the next section, we use this program to explore the circuit's behavior in various
changing environments through stochastic simulation.

\subsection*{Embodied inference is constrained by the logic of protein production}

Inference as embodied in a circuit made of proteins has a number of
properties in common with the abstract notion of
inference used in our analysis of a Markov model (Fig.~\ref{figMarkovModel}), but also some important
differences. We illustrate these properties below using stochastic
simulations of the circuit in different environments.

\begin{figure}[h]
\centering
\includegraphics{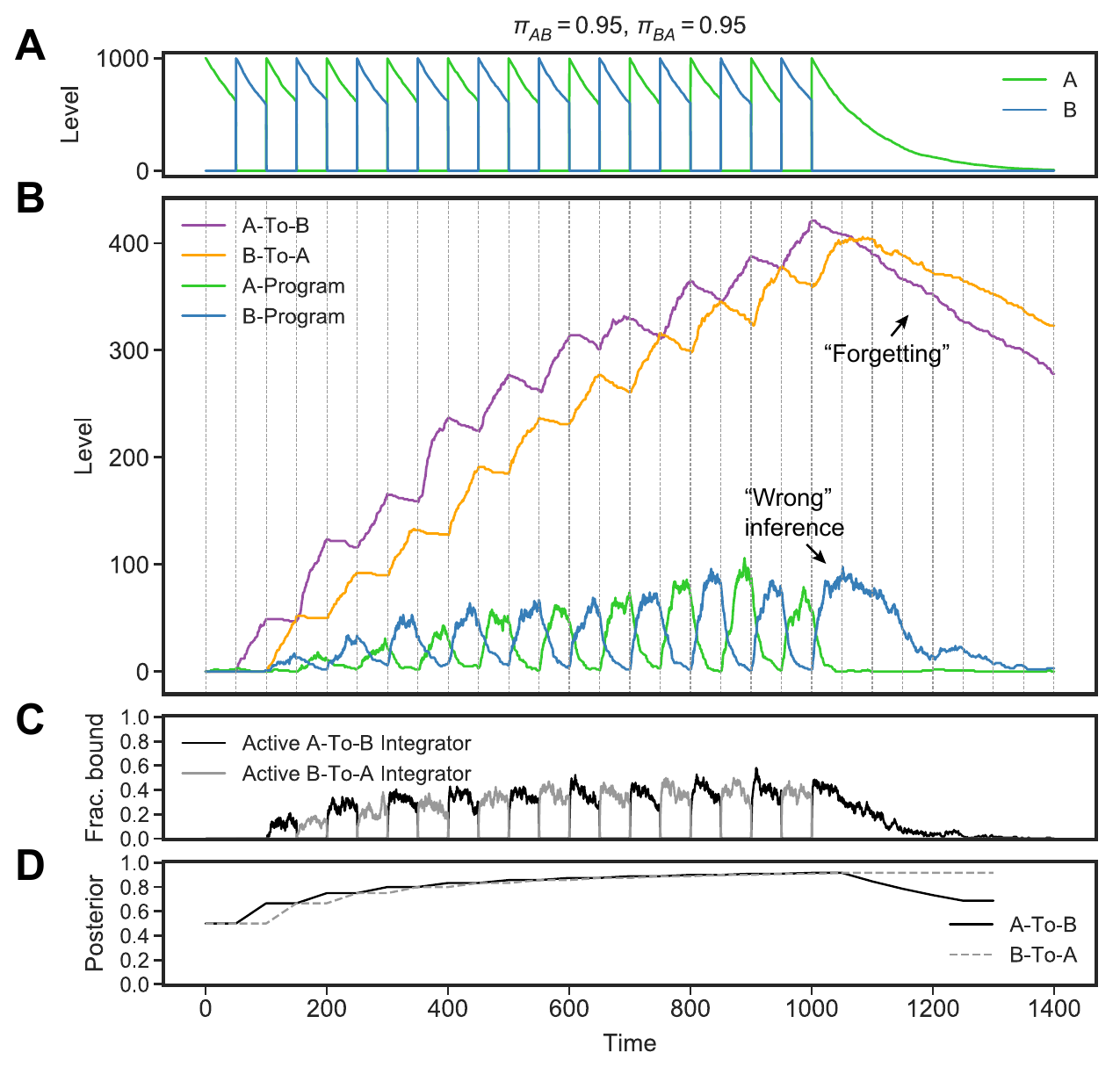}
\caption{Anticipatory circuit behavior. (A) Periodic environment that switches
between $A$ and $B$ (which are also degraded at some rate). (B) Levels
of circuit components (simulated in Kappa). The memory molecules $\AToB$ and $\BToA$ track
$A$ to $B$ and $B$ to $A$ transitions, respectively. As ``data''
accumulates, the molecular program associated with either the $A$ or
$B$ state gets upregulated in advance of switch. During last $A$
pulse, the $B$-associated program is ``wrongly''
upregulated. Over time, memory molecules degrade (``forgetting''). (C)
Fraction of $\AToB$ Integrator (black) and $\AToB$ Integrator (grey)
that are active. (D) Medians of posterior distribution over the transition probabilities $\ParamAB$ (black) and $\ParamBA$
(grey) at each time point, obtained from a discrete-time Markov model (with each data point
corresponding to a 50 time-unit interval) with a uniform prior.}
\label{figAnticipation}
\end{figure}

\begin{figure}[h]
\centering
\includegraphics{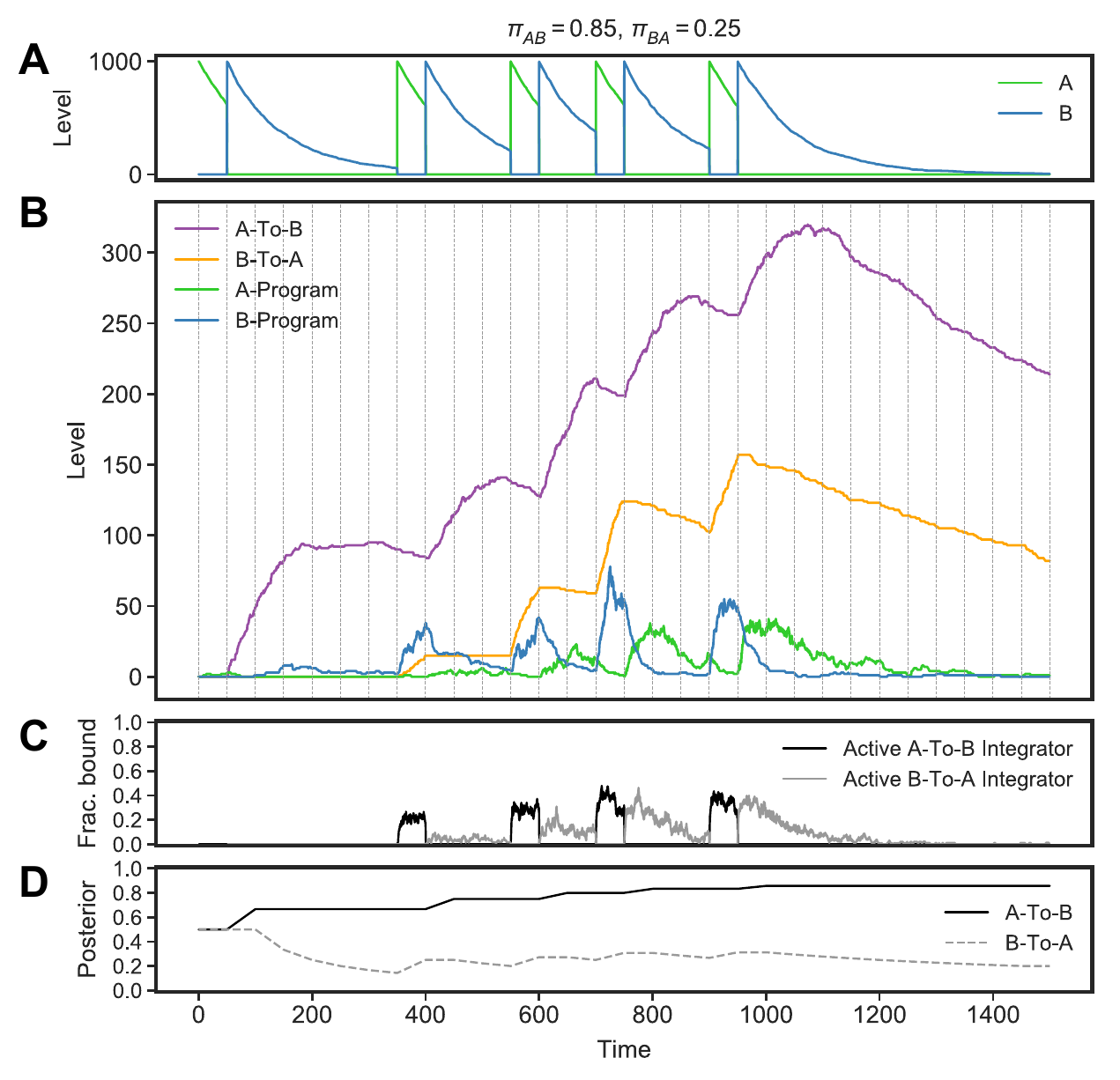}
\caption{Circuit behavior in an environment with asymmetric transition dynamics. (A)
  Environment as in Fig.~\ref{figAnticipation}, except $\ParamAB =
  0.85, \ParamBA = 0.25$ (an environment where $A$ is typically followed by longer sequence of
  $B$s). (B) Levels
of circuit components (simulated in Kappa). (C)
Fraction of $\AToB$ Integrator (black) and $\AToB$ Integrator (grey)
that are active. (D) Medians of posterior distribution over the transition probabilities $\ParamAB$ (black) and $\ParamBA$
(grey) at each time point using a discrete-time Markov model (with each data point
corresponding to a 50 time-unit interval) with a uniform prior.}
\label{figBehavior1}
\end{figure}

\begin{figure}[h]
\centering
\includegraphics{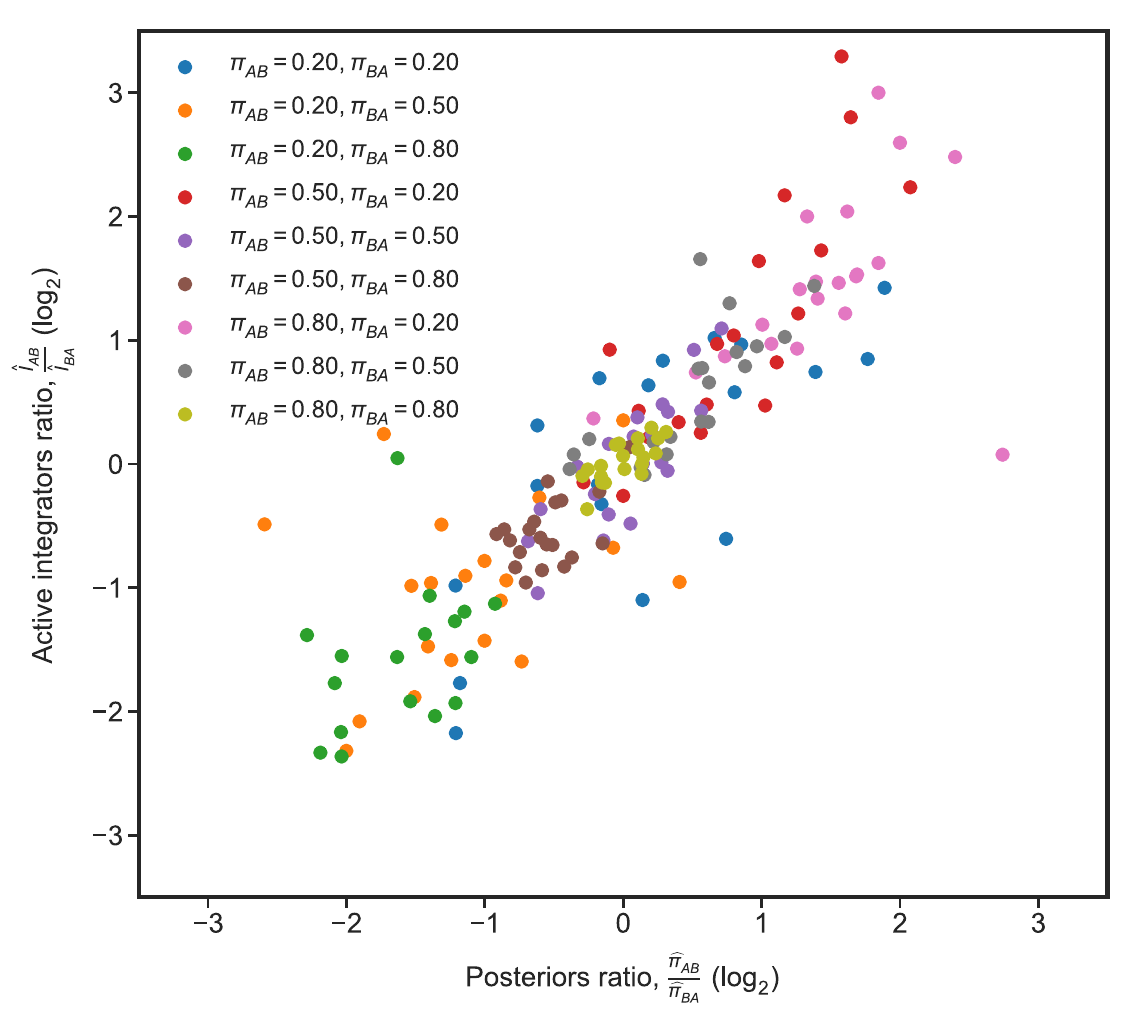}
\caption{Comparison of circuit behavior with idealized Bayesian
  posterior. Comparisons shown for 9 different configurations of
  $\ParamAB, \ParamBA$ that were used to produce changing environments
  consisting of 30 sequential perturbations (at 50 time-unit
  intervals). Each configuration comparison is plotted in a different
  color (20 simulations per configuration). On x-axis, ratio of
  posteriors for $\ParamAB$ and $\ParamBA$. $\EstParamAB$ is the median
  of the posterior over $\ParamAB$ through time (same for $\EstParamBA$). On y-axis, ratio of the
  median fraction (through time) of active $\AToB$ Integrator to that
  of active $\BToA$ Integrator. $\EstIntAB$ is median fraction of
  active $\AToB$ Integrator through time (same for $\EstIntBA$).}
\label{figPosteriorComparison}
\end{figure}

\textbf{Properties of an inferential mechanism.} The circuit
constructed by our procedure has several features that would be
expected of a mechanism that learns from its environment's history. These can
be illustrated in an environment that switches periodically from $A$
to $B$ at 50 time-unit intervals, as shown in
Fig.~\ref{figAnticipation}A. During the initial exposures to switches
(time $0-150$, Fig.~\ref{figAnticipation}B), there is little to no
change in $\AProgram$ and $\BProgram$ since the
memory molecules have not yet accumulated. As the circuit experiences
more ``data,'' an anticipation behavior emerges where $\BProgram$ is
upregulated ahead of the onset of $B$ (while still in $A$) and $\AProgram$ is
upregulated in advance of the onset of $B$. Similarly, $\BProgram$ begins to
be downregulated, in anticipation of $A$, while still in $B$ (and likewise for $\AProgram$). 

With more experiences of periodic switches, the
strength of the circuit's anticipatory behavior increases, consistent with the
idea that more experiences result in more confident estimates of
transition probabilities (mechanistically, this manifests in higher
concentrations of memory molecules; Fig.~\ref{figAnticipation}B). Furthermore,
since slightly more $A \rightarrow B$ switches were experienced than
$B \rightarrow A$ switches in this environment (Fig.~\ref{figAnticipation}A), the strength of activation for $\BProgram$
in the $A$ state is greater than that of the activation of $\AProgram$
in the $B$ state. The circuit's periodic behavior, which roughly
matches the 50 time-unit interval of switches in the environment, was not hardcoded in the
circuit but rather induced by the pattern of experience. When the periodic
pattern of switches in the environment is broken, and $A$ is not followed by $B$, the circuit
makes the ``wrong'' inference and still upregulates $\BProgram$
(Fig.~\ref{figAnticipation}B, time $1000$). After switching has stopped, the environment's imprint on the
circuit is then ``forgotten'' as the memory molecules degrade
(Fig.~\ref{figAnticipation}B, starting at time $1050$).

\textbf{Coupling of inference and action.} Through time, the fraction of activated integrator complexes reflect
the posterior over transition probabilities (Fig.~\ref{figAnticipation}C). Note that these glimpses
of the posterior are only available when the program of interest
($\AProgram$ and $\BProgram$) is being regulated. There is no
steadily available correlate of the posterior
probabilities. We compared this transient ``molecular correlate'' of the
posterior with the estimate of the posterior
over the transition probabilities ($\ParamAB$ and $\ParamBA$) based on a discrete-time two-state
Markov model with a uniform prior 
(treating $A$ and $B$ states as on/off and discretizing the
environment into 50 time-unit intervals;
Fig.~\ref{figAnticipation}D). Like the molecular
posterior of the circuit, the idealized estimate of $\ParamAB$ and
$\ParamBA$ increases with time as observations of the environment accumulate
(Fig.~\ref{figAnticipation}D).

\textbf{Constraints of embodiment by proteins.} The probabilistic model we used to
design the circuit was in discrete time, yet cells sense their
environment continuously. In the discrete-time model, by contrast, transitions from
a state to itself, e.g., $B \rightarrow B$, are recorded assuming a
fixed interval of time. Therefore, if $A$ tends to be followed by long
sequences of $B$, the estimate of $\ParamBA$ should decrease as the
count of $B \rightarrow B$ transitions increases
(Fig.~\ref{figBehavior1}). 

The logic of protein production captures this relationship without needing to
explicitly posit a fixed time interval. When an $A
\rightarrow B$ transition is experienced, the abundance of the $\AToB$ memory molecule
increases. If the environment remains in $B$, then the $\BToA$ memory
molecule, which we assume is degraded at some rate, will decrease in
abundance, and $B$ itself will degrade over time as well
(Fig.~\ref{figBehavior1}A,B). All things being equal, $\BToA$ will
decrease, making way for the repressor $\BToARep$ to act. As a
consequence, the fraction of active $\AToB$ Integrator
will be greater than the fraction of active $\BToA$ Integrator
(similarly to the idealized posterior, Fig.~\ref{figBehavior1}C,D). 

Note that the constraints of protein production also limit the
abstract probabilistic model we had started out with in other ways. While in the
Markov model the ``memory'' of the environment is determined by the
order of the model, here the extent of memory is bounded by the
degradation rates of the memory molecules. Without additional
mechanisms (such as positive feedback), the circuit's memory will be effectively
limited by the stability of the memory molecules themselves and by dilution (e.g., through cell division). 

\textbf{Variation in coupling inference to action.} In designing the circuit, we have not taken into account any special
considerations regarding how the state-specific expression programs
$\AProgram$ and $\BProgram$ should behave. It is easy to imagine that different contexts could impose specific constraints on the
relationship between the two programs and their kinetics. In some situations, it may be metabolically costly or toxic to have both $\AProgram$ and $\BProgram$
expressed at the same time. In others, anticipatory regulation of
one program may be metabolically cheaper than the other. In the current
design, the molecular estimate of the posterior, as realized by the fraction of activated integrators,
was used to regulate these programs without tuning the rate constants
so as to optimize for any particular constraint. Since continuous variation in rate constants (e.g., those affecting the
synthesis and degradation of $\AProgram$ and $\BProgram$) can produce
many different expression profiles, it is plausible that this aspect of
the circuit can be effectively shaped by selective forces. 

\textbf{Relationship to idealized inference.} As discussed above,
there is not a single quantity that serves as a
``molecular correlate'' of the posterior stably through time. The
closest quantity is the fraction of bound integrators, which is
coupled to action as described above. It is nonetheless instructive to compare this
correlate more systematically to the idealized posterior. To do this,
we simulated changing environments consisting produced using different
transition probabilities (Fig.~\ref{figPosteriorComparison}). For each
environment, we first estimated the median of the posterior over
$\ParamAB$ at each time point and then took $\EstParamAB$ to be the
median of those estimates (and similarly for $\ParamBA$ to get
$\EstParamBA$). We then computed a similar quantity for our circuit by 
taking the median of the fraction of bound $\AToB$ integrators through
time ($\EstIntAB$) and likewise for $\BToA$ integrators to get
$\EstIntBA$. Since these quantities are not directly comparable, we
compared the ratio of posteriors from the idealized model
($\frac{\EstParamAB}{\EstParamBA}$) with the ratio of active integrators in the circuit
($\frac{\EstIntAB}{\EstIntBA}$). The comparison shows broad agreement
between these quantities along different values of transition
probabilities (Fig.~\ref{figPosteriorComparison}), with the caveat
that both quantities are aggregated over time and thus do not capture
dynamic differences. 

\subsection*{Circuit failure modes}
\label{circuitFailures}

\begin{figure}[h]
\centering
\includegraphics{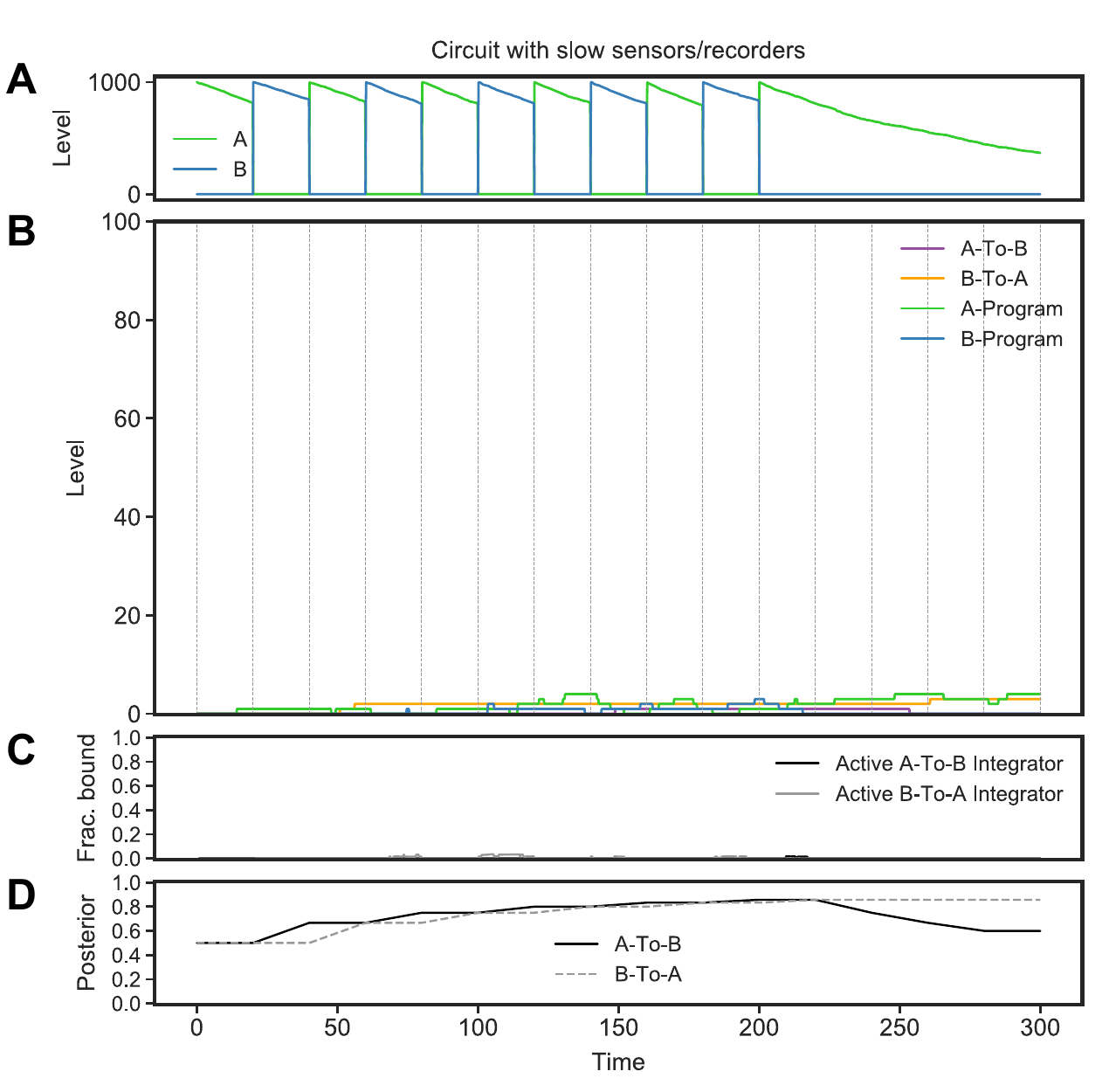}
\caption{Circuit with slow sensors/recorders (relative to timescale of
  switches in environment). Rate constants for sensors and recorders were
  set to be 100-fold slower than in Fig.~\ref{figAnticipation}. Periodic environment that switches every
  20 time-units. (A)-(D) as in Fig.~\ref{figAnticipation}.}
\label{figSlowCircuit}
\end{figure}

\begin{figure}[h]
\centering
\includegraphics{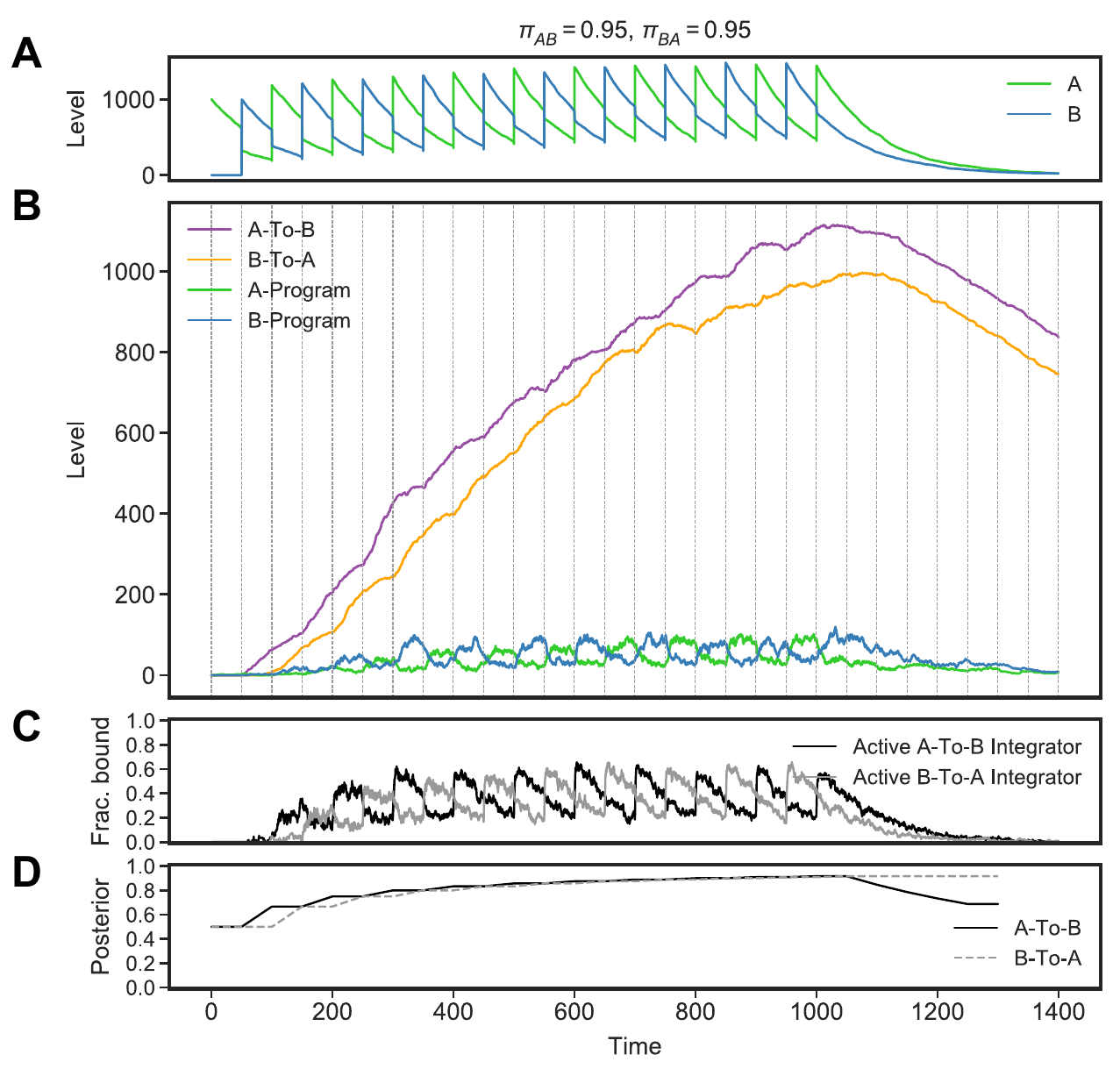}
\caption{Environment where only free (unbound) $A$ and $B$ are
  removed during switches. (A)-(D) as in Fig.~\ref{figAnticipation}.}
\label{figFreeCircuit}
\end{figure}

The circuit we have constructed only works when the dynamics of its
components are well-matched to the timescales of the environment's changes. When recorders and sensors are too slow relative to
duration of environment switches, for instance, the circuit fails to
record the environment's history (Fig.~\ref{figSlowCircuit}). More
subtle modes of failure arise from the circuit's sensitivity to the
abundance, not just the presence, of $A$ and $B$. While in the perturbations presented above we removed all
$A$ or $B$ molecules from the environment during each switch, in
practice, when the medium in which cells live is altered by an
experimenter, only the free (unbound) molecules are removed. When we
simulate an environment where only free molecules are removed in each
switch, residual $A$ and $B$ molecules accumulate with every switch
(Fig.~\ref{figFreeCircuit}). The circuit's behavior in this case is qualitatively similar to
the total perturbations (compare with Fig.~\ref{figAnticipation}), but
the circuit's response loses its punctate character (leading to
overall higher levels of $A$- and $B$-specific expression programs).

\begin{figure}[h]
\centering
\includegraphics{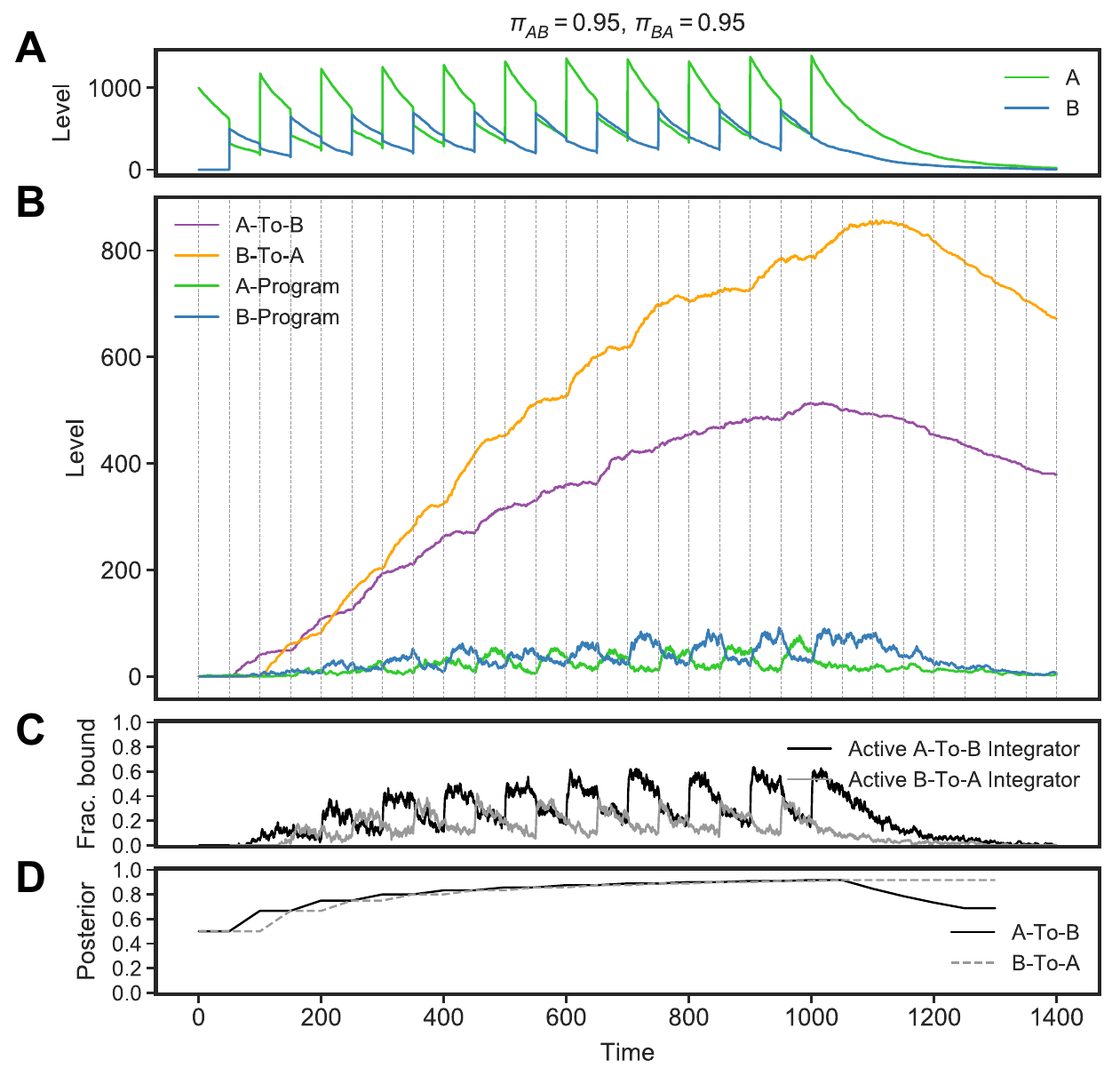}
\caption{Environment with variable perturbation amplitudes. 1000 units
  of $A$ are added during an $A$ addition compared to 500 of $B$ 
  in a $B$ addition (only free
  $A$ and $B$ are removed during switches). (A)-(D) as in Fig.~\ref{figAnticipation}.}
\label{figVariableCircuit}
\end{figure}

Since the circuit is sensitive to the abundance of the
signals and not just to their presence, pulses of varying amplitudes will
also affect the levels of the memory molecules. To examine this
behavior, we simulated an environment where $A$ additions were twice
the size of $B$ additions and only free $A$ or $B$ were removed
during switches (Fig.~\ref{figVariableCircuit}). This asymmetry in
pulse amplitude results in the circuit having an incorrect
representation of the environment's history. The higher abundance of $A$
results in substantially higher levels of $\BToA$ relative to $\AToB$, yet the number of $A \rightarrow
B$ switches is greater than that of $B \rightarrow A$ switches. The
resulting representation of the environment is also inconsistent:
while $[\BToA] \gg [\AToB]$, the bigger $A$ pulses---which bind the
$\AToB$ memory molecule---result in the
``posterior'' of $\ParamAB$ (i.e., fraction of activated $\AToB$
integrator) being consistently greater than the equivalent quantity for
$\ParamBA$. 

\subsection*{Searching for the molecular correlates of inference}

We have presented a synthetic design for a circuit that performs
inference. But how would one ``reverse-engineer'' a natural circuit
that performed a similarly sophisticated computation? A line of work by neuroscientists and molecular biologists has pointed
to cases where measuring input-output relationships of a dynamic and stateful
system (whether biological \citep{Marr1982Vision} or synthetic \citep{Mel2000ThoughtExp,Lazebnik2002Radio,JonasKording2017}), without a sufficiently detailed theoretical framework in
which to interpret these measurements, can lead to potentially misleading descriptions of how the underlying system
works. In systems biology, a common approach to studying cellular
pathways (enabled in part by high-throughput measurement technologies developed in recent years) has been to quantitatively
measure a pathway's component of interest as a function of
a quantitatively varying environmental input. When multiple inputs are
varied along a sufficiently large dynamic range, this measurement can
reveal surprising relationships between inputs and
outputs. For example, \citep{EscalanteChongRatio2015} measured the level of
induction of the galactose metabolic pathway (Gal) in yeast cells
grown in different concentrations of glucose (which represses Gal) and galactose (which
activates Gal) and found that Gal induction depends on the ratio of
glucose to galactose. These experiments are premised on obtaining
steady-state-like conditions where the level of the sugars (glucose
and galactose) and measured signal are relatively stable.

\begin{figure}[h]
\centering
\includegraphics[scale=0.9]{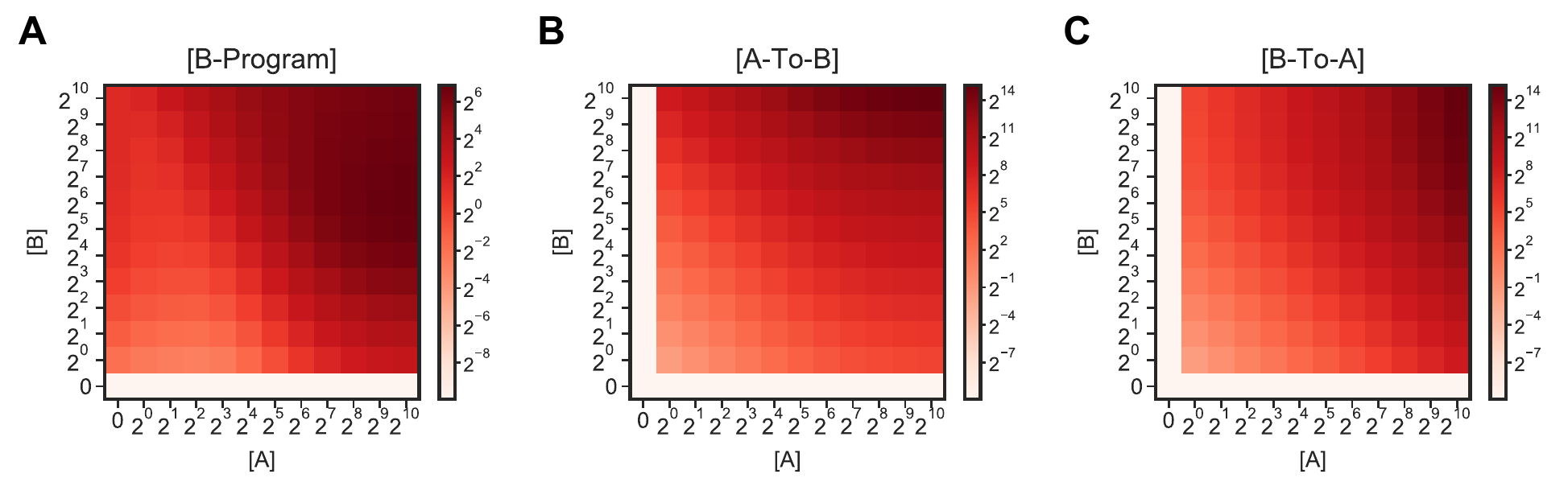}
\caption{Limitations of steady-state analysis. (A) Steady-state level of $B$-associated gene expression program
($\BProgram$). Estimated using deterministic simulation of ODEs corresponding
to circuit from Fig.~\ref{figCircuit} (initial levels of memory molecules,
$\AToB$ and $\BToA$, were set to $0$). (B) Steady-state level of $\AToB$
memory molecule. (C) Steady-state level of $\BToA$.}
\label{figDoubleGrad}
\end{figure}

Could a quantitative approach based on steady-states uncover inference
in our circuit? Using the ODE description of our circuit
(Fig.~\ref{figProgram}), we deterministically simulated the circuit across a range of concentrations of
the inputs ($A$ and $B$), which were held fixed, and examined the steady-state concentrations of $[\BProgram]$ and the two memory molecules,
$[\AToB]$ and $[\BToA]$ (Fig.~\ref{figDoubleGrad}). The resulting three profiles
are broadly similar. The $\BProgram$
profile suggests that the circuit is most active when $[A]$ and
$[B]$ are both high (and that $B$ must be present for $\BProgram$ to
be expressed) (Fig.~\ref{figDoubleGrad}A). The steady-state profiles of the memory molecules are
also relatively similar to each other across a broad range of
input concentrations (Fig.~\ref{figDoubleGrad}B,C).\footnote{The relatively small asymmetry between the
  $[\AToB]$ and $[\BToA]$ steady-state profiles, which occurs in
  the higher ranges of $[A]$ and $[B]$, is explained by the fixed
  level of the sensors ($\ASensor$ and $\BSensor$). Consider the case where $[A]$
  and $[B]$ are much higher than the concentrations of the
  sensors and $[B] \gg [A]$. In this case, $[A]$ will saturate the
  $\ASensor$ molecules, resulting in (at most) all of them being active. The
  activated sensors will then produce a potentially unbounded number of
  $\BRecorder$ molecules. Since $[B] \gg [A]$, the higher $[B]$ is,
  the more it can activate the pool of recorders to trigger production
  of $\AToB$. Therefore, when $[B] >
  [A]$, we expect $[\AToB] > [\BToA]$.} These measurements suggest that the
memory molecules are induced in proportion to $[A]$ and $[B]$, masking
the fact that each molecule's concentration encodes a distinct
direction of environmental change. This indicates that a standard steady-state
analysis, even if performed quantitatively, can obscure the logic
underlying a circuit that responds to the environment's history.

These observations suggest that it would be difficult to interpret the
circuit without having a theory about what it does. But what if we did
assume that this circuit performs probabilistic inference? Much work by cognitive scientists
and neuroscientists has used probabilistic inference as a framework
for explaining behavior, which prompted a search for the correlates of
these models in biological signals (see the growing ``Bayesian
brain'' literature, e.g., \citep{Knill2004BayesianBrain}). Given the
explanatory power of these models with respect to behavior, it
may seem reasonable to look for the molecular signatures and
associated ``knobs'' that correspond to the components of probabilistic
models, including the prior, the likelihood function and the
posterior. There are at least two conceptual difficulties that arise
in this pursuit that can be seen through the analysis of our circuit. First, while inference about (or perception of) the
environment is typically separated from ``action'' in these
computational models, in our circuit the two are intimately
coupled. As the time-dependent profiles of the circuit show, the closest
thing to a ``molecular correlate'' of a posterior-like quantity---the
fraction of activated integrators---only comes into being
\textit{through} action (regulation of a gene expression program such
as $\AProgram$ and $\BProgram$). There is no stable quantity that
serves as a correlate of the posterior through time.

A second difficulty with searching for the molecular basis of
inference is that the abstract components of the model may not map
neatly onto corresponding molecular knobs. Consider for instance the
question of \textit{where} in the circuit the ``prior'' lies. One way to conceive of the prior is as the propensity to change
one's prior belief from an initial state of having seen ``no
data'' to another belief state. Technically, the notion of an initial state
is artificial here since cells are continuously living inside their
environment and sensing its fluctuations. Setting this aside, if we assume that cells
start from some initial state (where $A$ and $B$ are
absent) then the propensity to change can depend on, among other
things, the initial concentration of the repressor molecules. An
alternative way to conceive of a ``prior'' here is as the amount of change
in the representation of the environment based on observation of
change; i.e., how much does influence one transition has on the
internal representation, such as our memory molecules. This notion of
``prior'' can also be shaped by multiple overlapping and not
necessarily modular features, such as the rate constants governing the 
interaction between the sensors, recorders and memory molecules. This
step can also be affected at the layer of ``action,'' for instance by
tweaking how sensitive the state-specific proteins $\AProgram$ and $\BProgram$ are to change in the levels of memory
molecules. Although the circuit is not a soup where all components
are enmeshed with all others, there still is not a clear knob one can
use to directly and selectively tune ``the prior.''

\section*{Discussion}
We began with an abstract computational problem: inference in a
Markov model. Following essentially Marr's
three-levels prescription \citep{Marr1982Vision}, we characterized
the computational task (level 1), derived the necessary
representation and algorithm for solving it (level 2), and used these
to guide the design of a biochemical realization (level 3). In this process, however, we ended up with a biochemical circuit that does not map to the
abstract computation we had started out with in the
way one might naively expect. The logic of protein production---i.e., the synthesis/degradation of proteins and
the kinetics of these processes---constrained and shaped the computational task we
had started with. 

\begin{figure}[t]
\centering
\includegraphics{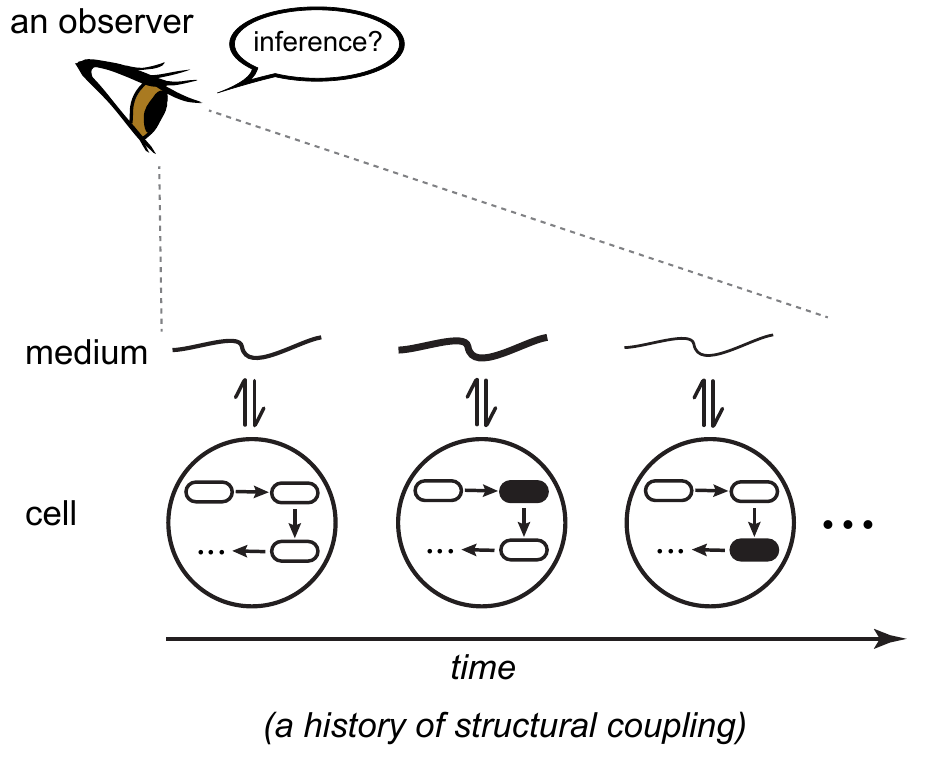}
\caption{Viewing cellular inference through the lens of a structural
  coupling. The cell interacts with a changing medium (squiggle line of
  varying thickness), which triggers changes in the cell's
  internal state (depicted by black and white rectangles). This structural coupling between cell and medium
  continues as long as the cell maintains its integrity. An observer,
  examining the history of this structural coupling, may posit that
  the cell is performing ``inference'' to anticipate its
  environment.}
\label{figObserver}
\end{figure}

Because it is embodied in proteins, our biochemical realization of
inference violates several dichotomies that are present in the
abstract computational characterization of inference (level 1). Specifically, in our
realization, there is no hard boundary
between environment versus cell, or between inference versus action. This can be framed along the lines of Maturana and Varela
\citep{MaturanaVarela1987Tree}, who conceive of organisms as \textit{structurally coupled} to their
environment. Perturbations generated outside the cell can push the cell's internal
state, while cells in turn shape the environment. This view does not
commit to sharp boundaries between an ``external'' environment and the
organism, or between the cell's perception of the environment and
the ``real'' state of the environment. To borrow from Merleau-Ponty \citep{MerleauPontyEyeAndMind}, the
environment awakens an echo in the cellular body, in our case through
the production of memory molecules (whose concentrations encode
probabilities that can be read out to realize inference). The circuit
we have presented in turn consumes the very signal that awakens the
echo, therefore shaping the environment and its perception. This picture is also inherently
dynamic: as long as the cell does not disintegrate, the structural
coupling between cell and environment continues and there is no need to invoke
steady-states (Fig.~\ref{figObserver}).\footnote{As Maturana and Varela put it: ``Ontogeny is
  the history of structural change in a unity without loss of
  organization in that unity. This ongoing structural change occurs in
  the unity from moment to moment, either as a change triggered by
  interactions coming from the environment in which it exists or as a
  result of its internal dynamics. As regards its continuous
  interactions with the environment, the cell unity classifies them
  and sees them in accordance with its structure at every
  instant. That structure, in turn, continuously changes because of
  its internal dynamics. The overall result is that the ontogenic
  transformation of a unity ceases only with its disintegration.''
  \citep[p.~74]{MaturanaVarela1987Tree}} This way of viewing cells
(and organisms) in relation to their environment is rooted in
approaches to cognition that emphasize embodied or enacted
perspectives (which have been applied to questions spanning artificial
intelligence, psychology, and biology \citep{WinogradFlores1986Understanding,MaturanaVarela1987Tree,ThelenSmith1996DynamicSystems,VarelaThompsonRosch2017EmbodiedMind}). 

Embodied perspectives come in tension with other approaches to cognition, particularly
when it concerns representations. In the ``rationalistic'' approach (see
\citep{WinogradFlores1986Understanding,Adam2006ArtificialKnowing} for
discussion) that is common within the fields of artificial intelligence and
cognitive science, representations play a central role. Organisms are
conceived as learning about the environment (which is generally
assumed to be external to the organism and often unaffected by it) by forming
representations of it. These representations are then used to guide
action through forms of information-processing operations over these representations (which
may be symbolic, non-symbolic, or a mixture). Our level 1 account of inference
exemplifies many of these assumptions common in rationalistic viewpoints.

Our attention here to the process by which a computational representation gets embodied can be seen as a compromise
between some rationalistic and embodied approaches. Our
work is not a triumph of embodied perspectives over
rationalistic ones, but rather a different framing of the relationship
between the two. The abstract rationalistic characterization was
indispensable in designing the circuit, in top-down fashion, and
arriving at (perhaps otherwise counterintuitive) constructions such as
the directional counter and the normalization-sampling circuit. The
computational framework (of level 1) is also valuable in making sense of the behavior of the resulting
circuit. Yet for all the reasons we outlined, there is not a clear-cut
mapping from the computational level of analysis to its ``hardware'' realization, as envisioned by
some rationalistic
accounts (including Marr's). The embodiment of the abstract
computation constrains and shapes the computation from level 1
in central ways that cannot be disregarded. 

The tension between the abstract notion of inference and
the biological condition of a cell in its environment can be further resolved by introducing
the notion of an observer. From the perspective of an observer, the
history of structural coupling between cell and its environment (or
medium) can be interpreted as a kind of ``inference''
(Fig.~\ref{figObserver}), but this construction lies in the observer's
cognitive domain.\footnote{Our view closely follows an
  argument made by Humberto Maturana, who pointed out that: ``From the perspective of
  the observer who considers an organism in a medium, the change of
  state that the perturbations by the medium trigger in it, may
  constitute representations of the medium, and the reactions of the
  organism that he beholds as a result of these perturbations may
  constitute indications, or descriptions, of the circumstances that
  constitute these representations. The observer, therefore, may
  describe the organism or its nervous system as acting, in their
  internally closed dynamics, upon representations of the medium while
  generating descriptions of it.'' \citep{Maturana1978Cognition}} Framing the interaction
between the levels as a constraining process---and keeping in mind
that ``inference'' is always attributed by an observer---may allow us
to throw out some of the rationalistic bathwater while keeping the
representational baby. 

Our view shares some aspects with the subsymbolic approach
to cognition (though without its commitments to specific neural
network mechanisms and related mathematical formalisms). The subsymbolic approach as formulated by
Smolensky \citep{Smolensky1986InfoProc}, for instance, sees the value
of a Marr-style top-down formulation of inference and acknowledges that inferential tasks can be
usefully described at multiple levels of abstraction. Importantly, however,
processes of implementation play a central role in subsymbolic theory,
which admits that the ``lower'' mechanistic level and the higher-level
descriptions mutually restrict each other and so in a way
have to be studied together. The same is true in our case. 

Numerous technical and conceptual questions about the biochemical
realization of inference are left. It remains to
be worked out how our procedure for generating biochemical circuits will scale to
more complex models---such as probabilistic models with numerous
(or latent) states---and what the metabolic properties of such
circuits would be. We have also not addressed the distinction between
inference at the level of individual cells versus populations, and more work is needed to
understand how the circuit would be affected by processes such as cell
division and growth. More fundamentally, it is unclear how such a
specialized circuit would get incorporated into an autopoietic
system \citep{MaturanaVarela1980Autopoiesis} where it would have to
work alongside general stress response and self-maintenance mechanisms. Finally, it would be worthwhile to examine how evolvable the circuit we have
presented is, following \citep{McGregor2012EvolutionAssociative} who evolved biochemical
circuits \textit{in silico} to implement an associative learning task (which resulted in
a circuit whose component abundances can also be correlated to posteriors in a
Bayesian model). It is possible that through simulated evolution, simpler or more
energetically efficient designs can be obtained (as argued by \citep{McGregor2012EvolutionAssociative}).

%


\section*{Acknowledgments}

We thank Andrew Bolton, Randy Gallistel and J\'{e}r\^{o}me Feret for helpful
comments on an earlier draft of this work.

\section*{Methods}
\label{Methods}

\subsection*{Real-time inference in a discrete-time Markov model}
We briefly derive a real-time procedure for inference in a Markov
model (which was used to generate the analytic estimate of the
posterior in all figures). We assume
a first-order discrete-time Markov model over two discrete states, $A$
and $B$, which is characterized by transition probabilities $\ParamAB$
and $\ParamBA$ (where $\ParamAA = 1 - \ParamAB$ and $\ParamBB = 1 - \ParamBA$). A sequence of observations is generated from the model by first
sampling an initial state $X_0$ (from a uniform prior over $A$ and $B$) and sampling subsequent
observations $X_1, X_2, X_3, \dots$ using the transition
probabilities: 

\begin{itemize}
\item If $X_t = A$, sample $X_{t+1} \sim
\Bern(1 - \ParamAB)$
\item Otherwise, if $X_t = B$, sample $X_{t+1}
\sim \Bern(\ParamBA)$. 
\end{itemize}

\noindent The quantity of interest here is the posterior distribution
over transition probabilities:
\begin{align*}
P(\ParamAB, \ParamBA \mid \history) &\propto \textrm{likelihood} \times
                            \textrm{prior}\\
&\propto P(\history
\mid \ParamAB, \ParamBA)P(\ParamAB, \ParamBA)
\end{align*}

\noindent where $\history$ stands for the sequence of observations of
past states, $\left<X_t, X_{t-1}, \dots\right>$. A standard approach is to use independent priors over the
transition probabilities, which lets us treat the posterior for each
parameter separately:
\begin{align*}
P(\ParamAB \mid \history) &\propto P(\history
\mid \ParamAB)P(\ParamAB)\\
P(\ParamBA \mid \history) &\propto P(\history
\mid \ParamBA)P(\ParamBA)
\end{align*} 

A convenient choice of prior over transition probabilities is the Beta distribution. A Beta distribution, $\BetaDist(\theta; \alpha_0,
\alpha_1)$, over $\theta \in [0, 1]$ is determined by a pair of
shape parameters $\alpha_0, \alpha_1$. Different settings of the alphas can be used to encode distinct beliefs about $\theta$. As examples: $\alpha_0 = \alpha_1 = 1$
is equivalent to a uniform distribution over $[0, 1]$, $\alpha_0 = \alpha_1 = 100$
is a unimodal distribution whose mode is at $\theta = 0.5$, and
$\alpha_0 < 1, \alpha_1 < 1$ (e.g., $\alpha_0 = \alpha_1 = 0.5$) is a U-shaped distribution corresponding to the belief that $\theta$
is likely to be close to 0 or 1. The Beta distribution is conjugate to the binomial
distribution \citep{GelmanBDA1995}, meaning that a
posterior that is the product of a Beta prior and a binomial
likelihood equals another Beta distribution whose parameters can be
calculated exactly. 

The observations of the
environment, $\history$, can be compressed into as
a set of counts (as described in main text) that are binomially
distributed, which allows us to use the Beta-Binomial conjugacy to
calculate our posteriors of interest analytically. For a two-state Markov model, the relevant counts are: (1)
$c_{AB}$, the number of $A \rightarrow B$ transitions, (2) the number of $B \rightarrow A$
transitions, and (3) the number of total switches $s$ in $\history$. (Note
that the other entries of the transition counts matrix $T$ can be
computed from this pair of summary statistics and $s$: $c_{AA} = s
- c_{AB}$ and $c_{BB} = s - c_{BA}$.) We can then expand the posterior over $\ParamAB$ as follows:
\begin{align*}
P(\ParamAB \mid \history) &\propto P(\history
                            \mid \ParamAB)P(\ParamAB)\\
&\propto P(c_{AB}, s
                             \mid \ParamAB)P(\ParamAB)\\
&\propto \Bino(C_{AB}; \ParamAB, s)\BetaDist(\ParamAB; \alpha_0,
  \alpha_1)\\
&= \BetaDist(\ParamAB; C_{AB} + \alpha_0, s - C_{AB} + \alpha_1)
\end{align*}

\noindent (and similarly for $\ParamBA$.)

This posterior can be computed in real-time up through the $k$th
observation (a task which is called ``filtering''
\citep{SarkkaBayesFilteringBook2013}) using a recursive
procedure that proceeds as follows. Let $C^{k}_{AB}$ be the number of $A \rightarrow B$
transitions up to the $k$th observation, and let $b^{k}_{\ParamAB}$ be the
posterior up to the $k$th observation. The base case of the procedure
is the first observation, which is the switch from $X_0
\rightarrow X_1$ (where $k = 1$):
\begin{align*}
b^{1}_{\ParamAB} &= P(\ParamAB \mid \left<X_1, X_0\right>) \\
 &\propto \Bino(C^{k}_{AB}; \ParamAB, k)\BetaDist(\ParamAB; \alpha_0,
  \alpha_1)\\
&= \BetaDist(\ParamAB; C^{k}_{AB} + \alpha_0, s - C_{AB} + \alpha_1)
\end{align*}

We can now define the $b^{k+1}_{\ParamAB}$ in terms of
$b^{k}_{\ParamAB}$ by noting that the posterior as of the $k$th observation,
$b^{k}_{\ParamAB}$, serves as the prior at $k+1$. Let $\alpha^{k}_0,
\alpha^{k}_1$ be the parameters of $b^{k}_{\ParamAB}$. Now let $I_{AB}$ be an
indicator variable that is 1 if $X_{k} \rightarrow X_{k+1}$ is an $A
\rightarrow B$ transition (and 0 otherwise), and $I_{AA}$ an indicator variable
that is 1 if $X_{k} \rightarrow X_{k+1}$ is an $A \rightarrow A$
transition (and 0 otherwise). Then the posterior after the $k+1$
observation is:
\begin{align*}
b^{k+1}_{\ParamAB} = \BetaDist(\ParamAB; \alpha^{k}_0 + I_{AB}, \alpha^{k}_1 + I_{AA})
\end{align*}

Note that if $X_{k} = B$, then $I_{AB} = I_{AA} = 0$, which means the estimate
of the posterior over $\ParamAB$ does not change (i.e., $b^{k+1}_{AB} = b^{k}_{AB}$). The
same derivation as above applies to the posterior over $\ParamBA$. 

To estimate the posterior over the transition probabilities in the above figures (e.g., Fig.~\ref{figAnticipation}D), we used
the median of the posterior distribution at each time step.

\subsection*{Stochastic and deterministic simulations of circuit}

Stochastic simulations were performed using Kappa's KaSim
(version: v4.0rc1-155-g93b4c2f). ODEs (in SBML
format) corresponding to the Kappa program were generated using the
KaDE utility \citep{Camporesi2017KaDE} and simulated using the
roadrunner library in Python \citep{Somogyi2015Roadrunner}. For
steady-state simulations in the presence of fixed inputs (Fig.~\ref{figDoubleGrad}), we ran the
ODE simulation until the sum of changes (across all species) between time bins was less
than a fixed threshold ($0.001$).

\begin{itemize}
\item
  \href{https://gist.github.com/yarden/959c1253c2eca61f3080d50a239a9b9c}{Programmatically
    generated Kappa program corresponding to two-state inference circuit}.
\item \href{https://gist.github.com/yarden/b233fa9ab8ca963ba34eb92d73f790b8}{ODEs corresponding to circuit} (in SBML format) generated
  by KaDE.
\end{itemize}

\FloatBarrier

\setlength{\bibsep}{0.0pt}

\bibliographystyle{apalike}
\bibliography{refs}      

\end{document}